%% file: main.tex
\documentclass[sigconf, nonacm]{acmart}
\usepackage{subcaption}
\usepackage{enumitem}
\usepackage{pgfplots}
\newcommand\vldbdoi{XX.XX/XXX.XX}
\newcommand\vldbpages{XXX-XXX}
\newcommand\vldbvolume{14}
\newcommand\vldbissue{1}
\newcommand\vldbyear{2020}
\newcommand\vldbauthors{\authors}
\newcommand\vldbtitle{\shorttitle} 
\newcommand\vldbavailabilityurl{https://github.com/uYanJX/QCFuse-Demo}
\newcommand\vldbpagestyle{plain} 

\usepackage{enumitem}
\setlist[itemize]{noitemsep, left=0pt} 
\usepackage{xcolor} 

\usepackage{subcaption}
\usepackage{stfloats}

\begin{document}
\title{QCFuse: Query-Centric Cache Fusion for Efficient RAG Inference}
\author{Jianxin Yan}
\affiliation{%
  \institution{Zhejiang University}
  \city{Hangzhou}
  \state{China}
}
\email{22521296@zju.edu.cn}

\author{Zeheng Qian}
\affiliation{%
  \institution{The University of Sydney}
  \streetaddress{}
  \city{Sydney}
  \country{Australia}
}
\email{zqia0047@uni.sydney.edu.au}

\author{Wangze Ni}
\affiliation{%
  \institution{Zhejiang University}
  \city{Hangzhou}
  \state{China}
}
\email{niwangze@zju.edu.cn}

\author{Zhitao Shen}
\affiliation{%
  \institution{Ant Group}
  \city{Shanghai}
  \state{Cina}
}
\email{zhitao.szt@antgroup.com}

\author{Zhiping Wang}
\affiliation{%
  \institution{Ant Group}
  \city{Shanghai}
  \state{China}
}
\email{laoman.wzp@antgroup.com}

\author{Haoyang Li}
\affiliation{%
  \institution{PolyU}
  \city{Hong Kong}
  \country{China}
}
\email{haoyang-comp.li@polyu.edu.hk}

\author{Jia Zhu}
\affiliation{%
  \institution{Zhejiang Normal University}
  \city{Jinhua}
  \country{China}
}
\email{jiazhu@zjnu.edu.cn}

\author{Lei Chen}
\affiliation{%
  \institution{HKUST(GZ) \& HKUST}
  \city{Guangzhou}
  \country{China}
}
\email{leichen@cse.ust.hk}

\author{Kui Ren}
\affiliation{%
  \institution{Zhejiang University}
  \city{Hangzhou}
  \state{China}
}
\email{kuiren@zju.edu.cn}


\input{01_abstract}
\maketitle

\pagestyle{\vldbpagestyle}
\begingroup\small\noindent\raggedright\textbf{PVLDB Reference Format:}\\
\vldbauthors. \vldbtitle. PVLDB, \vldbvolume(\vldbissue): \vldbpages, \vldbyear.\\
\href{https://doi.org/\vldbdoi}{doi:\vldbdoi}
\endgroup
\begingroup
\renewcommand\thefootnote{}\footnote{\noindent
This work is licensed under the Creative Commons BY-NC-ND 4.0 International License. Visit \url{https://creativecommons.org/licenses/by-nc-nd/4.0/} to view a copy of this license. For any use beyond those covered by this license, obtain permission by emailing \href{mailto:info@vldb.org}{info@vldb.org}. Copyright is held by the owner/author(s). Publication rights licensed to the VLDB Endowment. \\
\raggedright Proceedings of the VLDB Endowment, Vol. \vldbvolume, No. \vldbissue\ %
ISSN 2150-8097. \\
\href{https://doi.org/\vldbdoi}{doi:\vldbdoi} \\
}\addtocounter{footnote}{-1}\endgroup

\ifdefempty{\vldbavailabilityurl}{}{
\vspace{.3cm}
\begingroup\small\noindent\raggedright\textbf{PVLDB Artifact Availability:}\\
The source code, data, and/or other artifacts have been made available at \url{\vldbavailabilityurl}.
\endgroup
}

\section{Introduction}
\input{02_introduction}
\section{Related Work}
\input{03_related}
\section{System}
\input{04_system}
\section{Demonstration}
\input{05_demonstration}
\section{Conclusion}
\input{06_conclusion}





\bibliographystyle{ACM-Reference-Format}
\bibliography{sample}

\end{document}

%% file: 01_abstract.tex
\begin{abstract}
Cache fusion accelerates generation process of LLMs equipped with RAG through KV caching and selective token recomputation, thereby reducing computational costs and improving efficiency. However, existing methods primarily rely on local perspectives for token selection and lack global awareness from the user query. Utilizing this global awareness is challenging due to the high cost of obtaining context-aware query representations and the strict pipeline constraints required for efficient attention analysis. Thus, this demonstration introduces QCFuse, an innovative KV cache fusion system centered on the user query. QCFuse leverages semantic summary anchors to enhance query representations and selectively recomputes query-related tokens to improve accuracy, updating tokens based on the attention distribution of the most critical Transformer layer to preserve the high efficiency of the pipeline structure. Evaluations on real-world datasets demonstrate that QCFuse significantly improves the response efficiency of LLMs by 40\% while maintaining equivalent accuracy compared to current methods. Additionally, in certain scenarios, QCFuse achieves an attention denoising effect that yields higher response accuracy, demonstrating substantial potential in the optimization of LLM inference.
\end{abstract}

%% file: 02_introduction.tex
LLMs equipped with RAG are standard for enterprise knowledge-base question answering and content generation, as they mitigate hallucinations and support real-time knowledge updates. In high-concurrency production environments, however, RAG generation remains heavily bottlenecked. Although context chunks retrieved for different queries can overlap by over 70\%, strict prefix-matching policies prevent traditional prefix caching from reusing them. LLMs are thus forced to fully prefill redundant contexts. As a result, time to first token (TTFT) grows quadratically with context length, wasting immense computational resources.

\begin{figure}[t]
    \centering
    \includegraphics[width=0.49\textwidth]{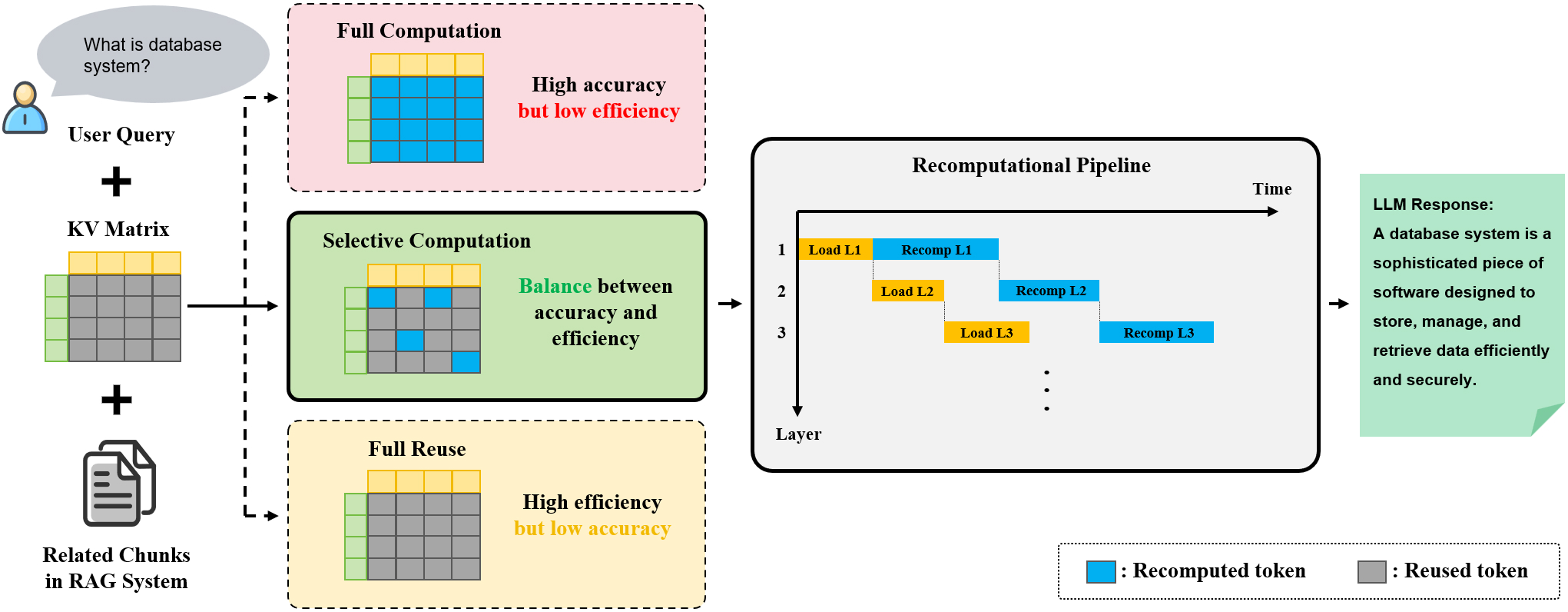}
    \caption{Comparison among Full Computation, Full Reuse, and Cache Fusion.} 
    \label{fig:compare}
    \vspace{-1.5em}
\end{figure}

As illustrated in Figure \ref{fig:compare}, cache fusion has emerged as a primary optimization strategy. These approaches merge historical KV caches and selectively recompute a subset of tokens. This reduces costs while preserving accuracy comparable to full computation. Prior work determines token recomputation targets in various ways: CacheBlend~\cite{Yao2024CacheBlendFL} uses numerical deviations of KV tokens, whereas EPIC~\cite{Hu2024EPICEP} statically recomputes a fixed ratio. Both methods effectively reduce computational overhead.

\begin{figure*}[t]
    \vspace{-4em}
    \centering
    \includegraphics[width=0.85\textwidth]{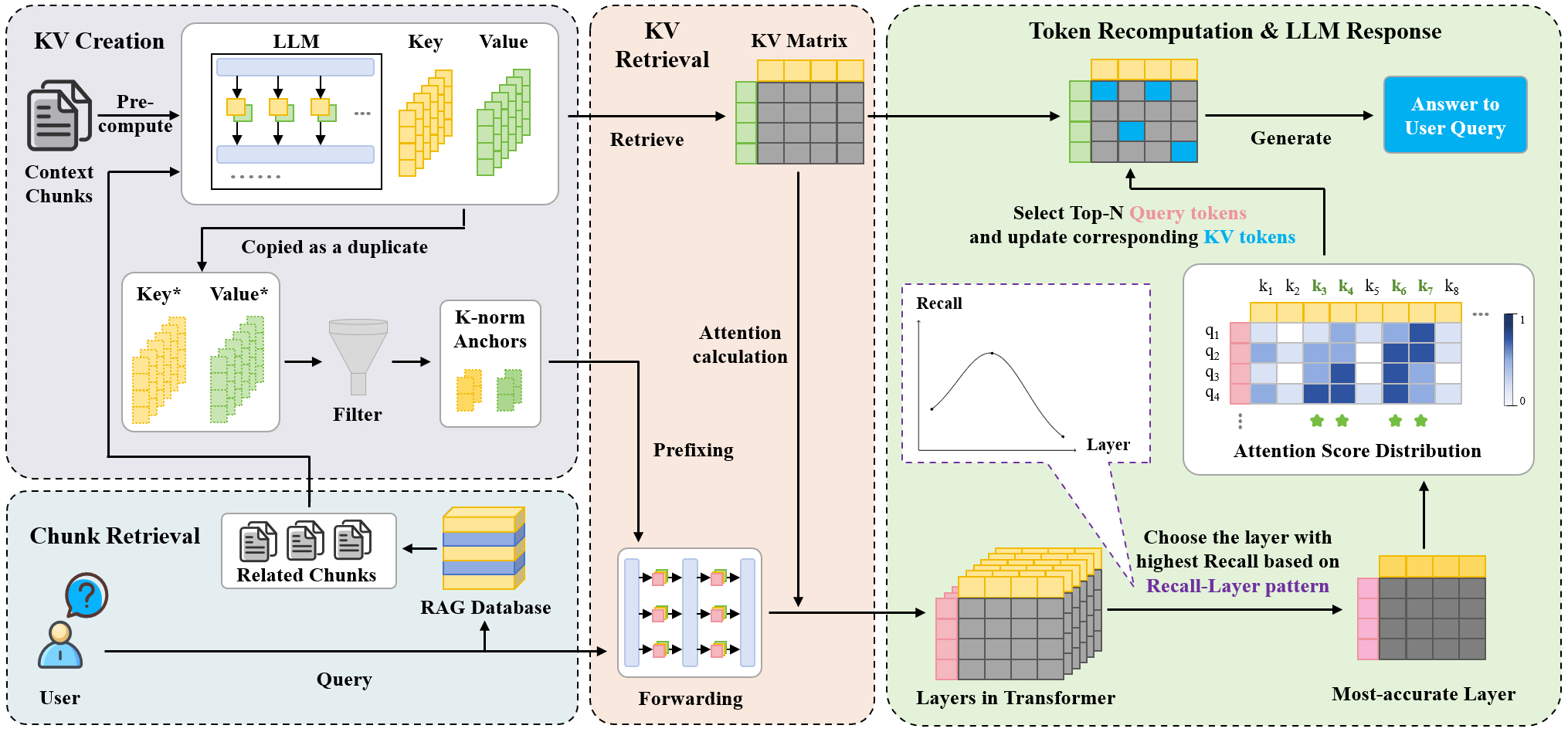}
    \caption{Architecture of the QCFuse System.} 
    \label{fig:process}
    \vspace{-1em}
\end{figure*}

A central limitation of existing methods is their lack of global awareness regarding the user query. By relying primarily on local cues, such as static positional heuristics or first-layer KV deviations, they overlook the query's role as the primary driver of the generation process. Operating on intermediate representations without considering the original request leads to suboptimal resource allocation. Computational budget is often spent on irrelevant tokens while critical ones are ignored, causing significant accuracy drops under aggressive acceleration.

Using the query's attention distribution over context tokens as a selection criterion is an intuitive alternative. Tokens with high query attention typically exert the greatest influence on generation quality. Realizing this within cache fusion systems, however, presents two primary challenges.
\begin{itemize}[topsep=0pt, itemsep=0pt, parsep=0pt]
\item \textbf{The first challenge concerns obtaining context-aware query representations at minimal cost}. A naive query-only forwarding process, which omits the contextual KV cache, yields ungrounded representations and unreliable attention distributions. Conversely, injecting the complete context KV cache during query forwarding disrupts the pipelined execution of cache fusion systems. To handle very long contexts, systems store full KV caches on SSDs and process them layer by layer. While the GPU recomputes one layer, the pipeline prefetches the next layer's KV from the SSD. Requiring the full context KV to be materialized beforehand forces the system to wait for all data to load, reducing the pipeline to sequential execution and eliminating efficiency gains. Acquiring context-enhanced query representations without disrupting this pipeline is thus essential.

\item \textbf{The second challenge involves efficient attention analysis within these pipeline constraints}. Computing query attention across all Transformer layers, as done in ProphetKV~\cite{Wang2026ProphetKVUS}, blocks the pipeline due to cross-layer dependencies. Relying solely on the final layer, as in FusionRAG~\cite{Wang2026FromPC}, provides an incomplete semantic view. The system must instead identify a single pipeline-friendly layer whose attention distribution serves as a reliable proxy for global token importance.
\end{itemize}

We address these challenges through two main technical contributions. \textbf{First, we propose anchor-based lightweight query probing.} By analyzing token key-norm magnitudes, we extract anchor tokens from each precomputed context chunk to serve as compressed semantic summaries. These tokens are injected as lightweight prefixes during query forwarding to produce context-enhanced query representations. Unlike previous context-free forwarding methods, this approach maintains pipeline efficiency.

\textbf{Second, we achieve semantic localization via critical-layer attention profiling.} Empirical findings suggest that middle layers offer superior semantic localization. We therefore analyze query attention at a single critical middle layer. This mechanism avoids the pipeline stalls associated with cross-layer dependencies and the incomplete views of last-layer approaches, successfully balancing accuracy and system efficiency.

Building on these mechanisms, this demonstration paper presents QCFuse, an efficient query-centric cache fusion system implemented on SGLang. The system features a high-performance sparse-attention Triton kernel for discrete token recomputation. Evaluations across multiple LLMs and multi-hop QA benchmarks demonstrate up to a 2$\times$ speedup of TTFT over full computation and a 40\% latency reduction compared to existing cache fusion baselines, with matching or improved generation quality. We demonstrate how QCFuse integrates seamlessly into enterprise knowledge assistants to deliver near-real-time answers over massive document collections.

%% file: 03_related.tex
\textbf{KV and Prefix Caching.} KV and prefix caching~\cite{Zheng2023SGLangEE} accelerates large language model inference by reusing historical Key/Value matrices, avoiding redundant computation and reducing latency. Frameworks such as SGLang~\cite{Zheng2023SGLangEE} implement this via prefix-tree matching. While highly effective for constant prefixes, these mechanisms struggle with the dynamic context assemblies typical of RAG applications, where retrieval chunks are frequently reordered or inserted into the middle of the prompt.

\noindent \textbf{Location-independent Caching and Selective Recomputation.} To enable cache reuse in dynamic contexts, fusion methods concatenate discrete KV chunks and selectively recompute tokens that experience semantic shifts. CacheBlend~\cite{Yao2024CacheBlendFL} identifies these tokens using first-layer KV deviations, EPIC~\cite{Hu2024EPICEP} defaults to a static ratio at the sequence start, and KVShare~\cite{Yang2025KVShareAL} scales KV deviation by initial attention weights. More recently, ProphetKV~\cite{Wang2026ProphetKVUS} and FusionRAG~\cite{Wang2026FromPC} proposed query-guided token selection, yet both encounter pipeline synchronization issues. ProphetKV evaluates all layers concurrently, ignoring stage-wise scheduling constraints. FusionRAG computes an initial query pass without context, relying on last-layer attention that provides insufficient semantic grounding. We emulate these two query-centric baselines in our evaluations as QCAll and QCLast.

%% file: 04_system.tex
\subsection{System Workflow}
As shown in Figure \ref{fig:process}, the workflow of QCFuse consists of four highly optimized phases:

\noindent \textbf{Phase 1: Offline Pre-computation and Anchor Extraction.} Before query processing, the system pre-computes the KV cache for all context chunks in the RAG database. During this stage, the full pre-computed KV cache is stored persistently on the SSD. Concurrently, a minor fraction of tokens with the highest key-norm values are copied and extracted to serve as a compressed semantic summary. Due to their minimal storage footprint, these anchor KV tokens are stored directly in the CPU memory to minimize latency. This offline procedure completely avoids any time overhead during the online generation process.

\noindent \textbf{Phase 2: RAG Retrieval and Context-aware Query Probing.} When a user query enters the scheduler of the SGLang, the system performs a standard RAG retrieval process to fetch relevant chunks. Instead of conducting a context-free query forwarding process, the system utilizes the CPU-resident KV token anchors corresponding to the retrieved chunks. These anchors are injected into the GPU as lightweight prefixes alongside the query, endowing the initial query representations with profound contextual grounding without initiating massive data transfers from the SSD.

\noindent \textbf{Phase 3: Critical-layer Attention Analysis.} Following the query forwarding, the system exclusively loads the key (K) cache of the most critical middle layer from the SSD. It then performs an attention analysis between the query (Q) cache of the user request and the K cache of this specific critical layer. The resulting attention weights dictate the Top-$N$ context tokens to which the query relates most strongly, thereby yielding the essential indices for token reconstruction.

\noindent \textbf{Phase 4: Pipelined Cache Reconstruction and Response Generation.} Guided by these Top-$N$ indices, the GPU initiates discrete token recomputation, which adheres to a strictly pipelined architecture: while the GPU reconstructs the selected tokens for layer $i$, the pipeline simultaneously prefetches the KV cache for layer $i+1$ from the SSD. Ultimately, the updated and contextually enriched matrix set of KV tokens is fed into the native decoding engine of the SGLang framework for response generation with exceptionally low latency.

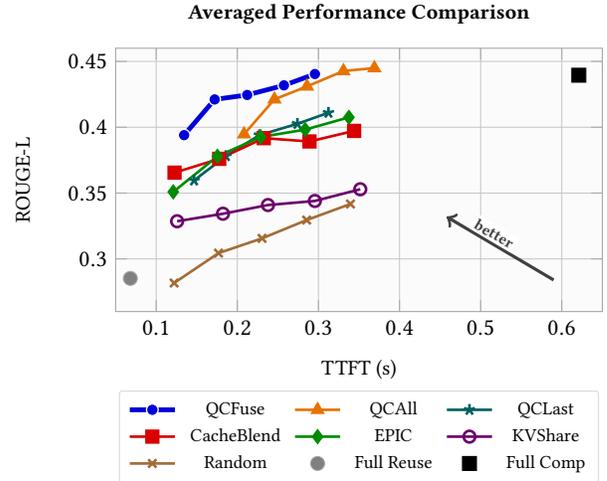
\begin{figure}[t]
\centering
\begin{tikzpicture}
\begin{axis}[
    xlabel={TTFT (s)},
    ylabel={ROUGE-L},
    title={Averaged Performance Comparison},
    title style={font=\small\bfseries},
    xlabel style={font=\small},
    ylabel style={font=\small},
    grid=both,
    major grid style={line width=.25pt,draw=gray!45},
    minor grid style={line width=.1pt,draw=gray!20},
    axis background/.style={fill=gray!3},
    axis line style={draw=gray!70},
    tick style={draw=gray!70},
    xmin=0.05, xmax=0.65,
    ymin=0.26, ymax=0.46,
    width=0.95\columnwidth,
    height=0.6\columnwidth,
    mark size=2.2pt,
    legend style={
        font=\footnotesize,
        draw=gray!30,
        fill=white,
        rounded corners=1pt,
        at={(0.5,-0.3)},
        anchor=north,
        legend columns=3,
        column sep=3pt,
        /tikz/every even column/.style={column sep=3pt}
    },
    cycle list name=exotic,
]

\addplot[
    mark=*,
    color=blue!85!black,
    mark options={fill=blue!85!black, draw=white, line width=0.5pt},
    line width=1.8pt,
    solid
]
coordinates {
    (0.1344, 0.3940)
    (0.1721, 0.4211)
    (0.2122, 0.4245)
    (0.2575, 0.4318)
    (0.2957, 0.4404)
};
\addlegendentry{QCFuse}

\addplot[
    mark=triangle*,
    color=orange!90!black,
    mark options={fill=orange!90!black},
    line width=1.0pt,
    solid
]
coordinates {
    (0.2083, 0.3947)
    (0.2458, 0.4212)
    (0.2859, 0.4310)
    (0.3309, 0.4427)
    (0.3690, 0.4449)
};
\addlegendentry{QCAll}

\addplot[
    mark=star,
    color=teal!75!black,
    mark options={fill=teal!75!black},
    line width=1.0pt,
    solid
]
coordinates {
    (0.1467, 0.3596)
    (0.1859, 0.3781)
    (0.2273, 0.3940)
    (0.2741, 0.4026)
    (0.3124, 0.4110)
};
\addlegendentry{QCLast}

\addplot[
    mark=square*,
    color=red!85!black,
    mark options={fill=red!85!black},
    line width=1.0pt,
    solid
]
coordinates {
    (0.1226, 0.3655)
    (0.1777, 0.3760)
    (0.2326, 0.3917)
    (0.2890, 0.3892)
    (0.3440, 0.3972)
};
\addlegendentry{CacheBlend}

\addplot[
    mark=diamond*,
    color=green!55!black,
    mark options={fill=green!55!black},
    line width=1.0pt,
    solid
]
coordinates {
    (0.1210, 0.3509)
    (0.1755, 0.3779)
    (0.2288, 0.3926)
    (0.2837, 0.3982)
    (0.3376, 0.4075)
};
\addlegendentry{EPIC}

\addplot[
    mark=o,
    color=violet!85!black,
    mark options={fill=violet!35},
    line width=1.0pt,
    solid
]
coordinates {
    (0.1258, 0.3286)
    (0.1823, 0.3343)
    (0.2378, 0.3409)
    (0.2956, 0.3440)
    (0.3516, 0.3529)
};
\addlegendentry{KVShare}

\addplot[
    mark=x,
    color=brown!85!black,
    mark options={solid},
    line width=1.0pt,
    solid
]
coordinates {
    (0.1221, 0.2817)
    (0.1768, 0.3044)
    (0.2305, 0.3156)
    (0.2855, 0.3295)
    (0.3395, 0.3417)
};
\addlegendentry{Random}

\addplot[
    only marks,
    mark=*,
    color=gray!70,
    mark options={fill=gray, scale=1.2}
]
coordinates {
    (0.0681, 0.2852)
};
\addlegendentry{Full Reuse}

\addplot[
    only marks,
    mark=square*,
    color=black,
    mark options={fill=black, scale=1.2}
]
coordinates {
    (0.6210, 0.4395)
};
\addlegendentry{Full Comp}

\draw[->, very thick, black!75]
    (rel axis cs:0.90,0.12) -- (rel axis cs:0.68,0.36)
    node[pos=0.6, above, sloped, fill=white, inner sep=1pt, font=\scriptsize\bfseries] {better};

\end{axis}
\end{tikzpicture}
\caption{Average ROUGE-L vs. TTFT of existing methods under different recomputation ratios (0.1--0.5), averaged across models and datasets.}
\label{fig:performance_comparison}
\vspace{-2em}
\end{figure}

\begin{figure*}[t]
    \vspace{-4em}
    \centering
    \includegraphics[width=0.7\textwidth]{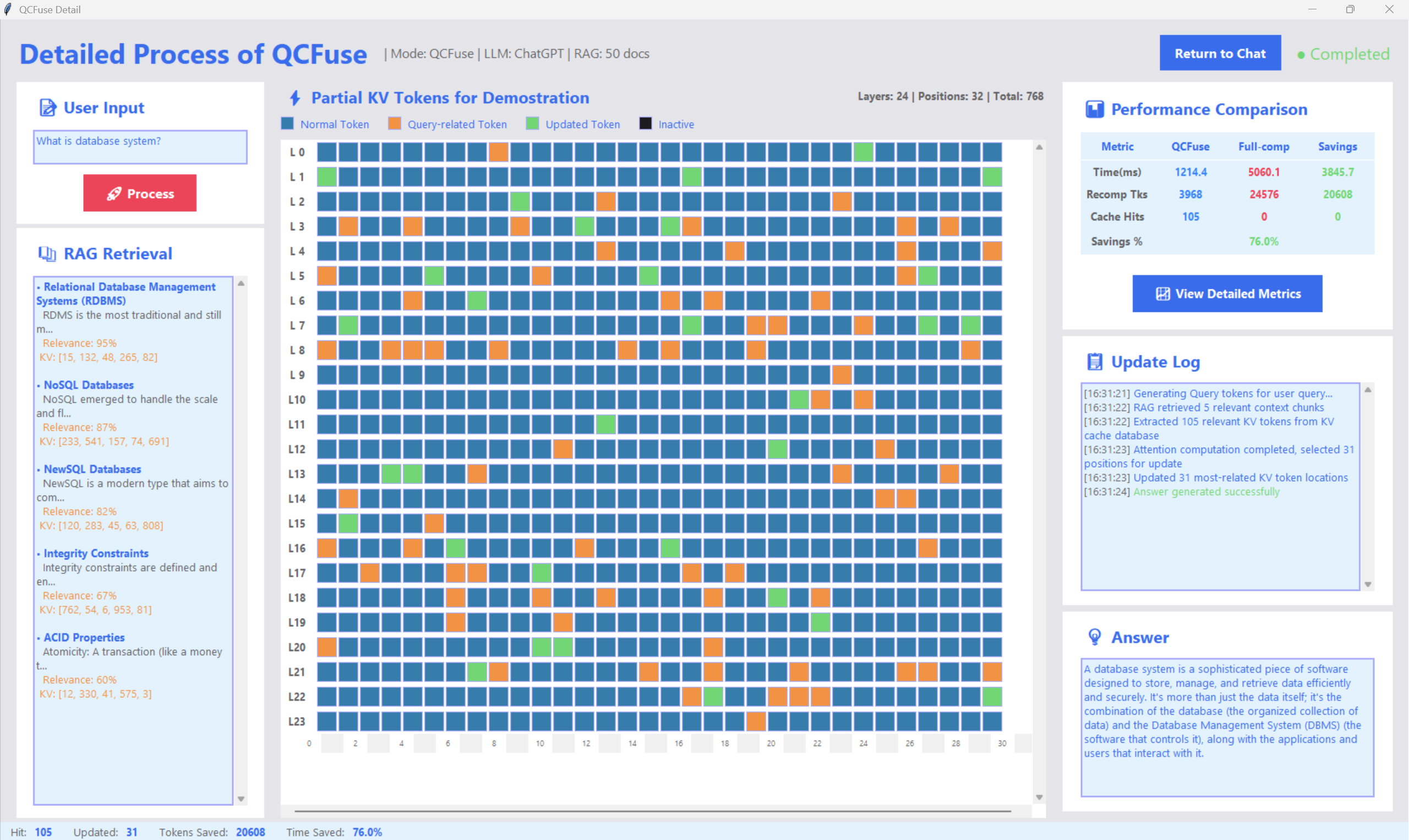}
    \caption{Detailed Interface of QCFuse for the Demonstration of KV Recomputation.} 
    \label{fig:demos2}
    \vspace{-1em}
\end{figure*}

\subsection{System Implementation}
To implement QCFuse, we modify the process of the SGLang framework at a minimal scale for the following key components:

\noindent \textbf{Location-independent Cache Indexing Module.} This module extends the native RadixCache of the SGLang framework to create a hash indexing table for context chunks. The KV tokens for each context chunk are independently cached. This configuration supports precise content-based hash searches while remaining intrinsically compatible with existing prefix caching logic.

\noindent \textbf{Query-related Token Selector.} This selector is executed during Phase 3. By forwarding the query with anchor prefixes and exclusively analyzing the attention of the critical middle layer, this module rapidly estimates the attention distribution of the query across all contextual tokens. After averaging and ranking the multi-head attention weights, the Top-$N$ tokens are formally selected as the corresponding set for recomputation. The recomputation ratio can be dynamically configured based on functional requirements for accuracy and performance.

\noindent \textbf{Location-aware Sparse Attention Kernel.} This kernel is a customized location-aware attention operator developed via Triton, which is completely compatible with the operator invocation interface of the SGLang framework. The kernel supports receiving a table of discrete token indices alongside their corresponding absolute locations to construct attention masks that faithfully follow causal constraints. This robust design guarantees absolute semantic correctness during the token recomputation process.

\subsection{Core Technique and Evaluation}
We evaluate QCFuse on one A100 GPU (80GB) with tested models including Llama3.1-8B, Qwen3-8B, and Mistral-v0.3-7B. We apply three question-answering datasets: Musique, 2WikiMQA, and HotpotQA, aiming to simulate real RAG features.

\noindent \textbf{Performance of the Query-related Global Assessing Strategy.} Our testing indicates that QCFuse achieves a ROUGE-L score 2.3 to 3.5 points higher than CacheBlend~\cite{Yao2024CacheBlendFL}. At a 40\% recomputation ratio, QCFuse matches the accuracy of full computation. On the HotpotQA dataset, QCFuse is 0.8 points better than full computation because it removes attention interactions with irrelevant tokens. This directly proves the attention denoising effect and improves the overall accuracy. Furthermore, QCFuse achieves accuracy comparable to QCAll with lower latency, while delivering substantially higher accuracy than QCLast.

\noindent \textbf{Effect of Sparse Attention Kernels.} Our sparse attention kernel only accelerates the partial computation phase. After applying this kernel across all tested schemes to ensure a fair comparison, as shown in Figure \ref{fig:performance_comparison}, QCFuse is two times faster than full computation. When QCFuse reaches the same accuracy as the baseline, it reduces the delay by an extra 40\%, which perfectly meets the strict requirements of fast RAG tasks.

%% file: 05_demonstration.tex
The demonstration system of QCFuse is developed using a front-backend separation architecture based on React and FastAPI. The frontend utilizes React-based charting libraries for dynamic visualizations, while the backend serves via the SGLang framework integrated with the QCFuse extension to facilitate real-time comparisons with baseline solutions. The demonstration interactively comprises two consecutive scenarios that progressively highlight the core advantages of the system, allowing users to input customized queries, adjust parameters, and verify results.

\noindent \textbf{Scenario 1: Conversational Interaction between User and LLMs.} This scenario showcases a standard chat interface of LLMs. However, before entering a query, the user can review all the context chunks stored in the RAG database, along with their summaries, hash index values, and other related information displayed on the left side of the interface. If no relevant chunks are available, the user can upload custom context files to the system. These files can be in standard contextual formats (e.g., .txt, .pdf, .csv) and are automatically pre-computed in real time. Subsequently, the user can select a preferred LLM and cache mode to execute the generation process of RAG. This process supports different LLMs and various cache modes, including QCFuse and other baselines such as CacheBlend, thereby facilitating a direct comparison of user experience between QCFuse and alternative methods.

\noindent \textbf{Scenario 2: Token Retrieval and Update.} 
Figure \ref{fig:demos2} demonstrates the core capabilities of QCFuse: independent chunk-level KV caching and location-independent token reuse. After receiving the response of the LLM on the chat interface, the user can click the "View Details" button at the bottom-right corner of each response bubble to inspect the details of the generation process.
\begin{itemize}[topsep=0pt, itemsep=0pt, parsep=0pt]
    \item Query-related context chunks retrieved from the RAG database are labeled with a content summary, relevance scores, and other information. These are exhibited on the left side of the interface as the base materials for the retrieval of KV tokens.
    \item After clicking the "Process" button, the user can observe changes in partially displayed KV tokens in the middle of the interface. The query-related KV tokens (i.e., those with high attention scores) are first uniformly highlighted in orange based on the critical layer. Subsequently, during the recomputation process, they are marked in green layer by layer as they are updated.
    \item The basic parameters (e.g., the number of computed tokens and the time cost) of the currently selected recomputation method and full computation are displayed on the right side of the interface, along with the process timeline of the system and the final answer generated by the LLM. The user can also click the "View Detailed Metrics" button to monitor other system parameters, such as cache storage and memory usage.
\end{itemize}

%% file: 06_conclusion.tex
We innovatively develop QCFuse, a query-centric KV cache fusion system that selectively recomputes query-related tokens. This system optimizes the generation process of LLMs equipped with RAG, achieving higher accuracy and superior efficiency.